\documentclass [12 pt]{article}
\usepackage{epsfig}
\newcommand {\spc} {\;\;\;}

\newcommand {\lbr} {\left [}
\newcommand {\rbr} {\right ]}

\newcommand {\uu} {$\_$}

\newcommand {\sig} {\sigma}
\begin{document}
\title{Theory of the friction force using electron cooling as an intrabeam
       scattering process}
\author{George Parzen}
\date{NOVEMBER 2006 \\BNL REPORT  C-A/AP NO.261}
\maketitle
\begin{abstract}
Using the results found previously [5] for the cooling rates of the
emittances, due to collisions between the electrons and the ions, a
result is found for the friction force acting on the ions. It is
shown that the friction force found here when used to track the ion
bunch will give the same emittance cooling rates as those found using
the intrabeam scattering theory for electron cooling [5].For the case
of the uniform in space
electron bunch distribution, the friction force found here agrees with the friction
force result found with the usual theory of electron cooling. 
\end{abstract}

\section {Introduction}
Using the results found previously [5] for the cooling rates of the
emittances, due to collisions between the electrons and the ions, a
result is found for the friction force acting on the ions. It is
shown that the friction force found here when used to track the ion
bunch will give the same emittance cooling rates as those found using
the intrabeam scattering theory for electron cooling [5].For the case
of the uniform in space
electron bunch distribution, the friction force found here agrees with the friction
force result found with the usual theory of electron cooling. 

\section{Intrabeam scattering, ions on ions}

Consider a beam which consists of a single bunch of completely ionized
ions.
The ions are doing betatron oscillations  in the transverse direction ,
and synchrotron oscillations in the longitudinal direction. In
addition the  ions are subject to the coulomb repulsion between
them. The scattering of each ion  by the other ions is called intrabeam
scattering. In Rhic , intrabeam scattering (IBS)  causes the beam
dimensions to grow slowly in all three diretions.The growth of the
beam can be computed using intrabeam scattering theory [1-4].

\section{Intrabeam scattering, ions on electrons}
In electron cooling the ion bunch is overlapped by an electron bunch which
is moving at the same velocity as the ion bunch. The ions can now
scatter off each other or they can scatter off the electrons in the
electron bunch.The scattering of the ions from each other occurs all
around the accelerator ring and causes the emittances of the beam to
grow. The scattering of the ions from the electrons occurs only in the
cooling section and causes negative growth (cooling) in the ion emittances.
Each kind of scattering may be considered as a kind of intrabeam scattering
and the growth of the ion bunch due to each kind of scattering can be
computed in the same way [5]. 

\section{Friction force definition using intrabeam scattering results 
for electron cooling}

The friction force will be defined as a force which acting on each ion
in the ion bunch will produce the same cooling rates, due to collisions
between ions and electrons, for the three
quantities, $<p_i^2>, \spc i=x,y,s$ for the ions as that found by the IBS theory
for electron cooling. $<p_i^2>$ is the average of $p_i^2$ over all the
ions in the bunch. It will be shown below that the friction force found using this
definition when used to track the ion bunch 
will give the same emittance cooling rates as those found using the
IBS theory of electron cooling.

\section{Friction force results found using intrabeam scattering results 
for electron cooling}

Using the results for the cooling rates of $<p_i^2>, \spc i=x,y,s$
found by the IBS theory for electron cooling  [5], and the above definition
of the friction force, one finds the following expression for the
friction force. The subscripts $a,b$ indicate ions and electrons
respectively. $N_b f_b(x,v_b)$ is the electron distribution
function. $N_b$ is the total nuber of electrons in the electron
bunch. $x$ is the location of the ion. The derivation of the friction
force results are given below. Using the Coulomb cross-section, one gets 
\begin{eqnarray}
\sigma_{ab} &=& (\frac {r_{ab}} {\beta_{ab}^2})^2 \frac{1}{(1-cos \theta)^2} 
   \spc \spc coulomb\; cross-section\;in\; CMS            \nonumber\\ 
r_{ab} &=& \frac{Z_aZ_b e^2}{\mu c^2}     \nonumber\\
\beta_{ab} c &=& |\vec{v_a}-\vec{v_b}|   \nonumber\\
\frac{1}{\mu} &=& \frac{1}{m_a}+\frac{1}{m_b}     \nonumber\\
    & &                    \nonumber\\
F_i &=& -4 \pi m_b N_b r_{ab}^2 c^4 \int d^3v_b \spc
     \frac{(v_a-v_b)_i}{|v_a-v_b|^3} 
  f_b(x,v_b) ln \lbr \frac{\beta_{ab}^2 b_{maxab}}{r_{ab}} \rbr
          \nonumber\\
\end{eqnarray}

One can also  find a result for any $\sig_{ab}$, and not just the
coulomb $\sig_{ab}$.
\begin{eqnarray*}
F_i &=& -2 \pi m_b N_b \int d^3v_b \spc \lbr (v_a-v_b)_i |v_a-v_b| f_b(x,v_b) 
     \int d\theta sin \theta (1-cos\theta) \sig_{ab} \rbr    \nonumber\\
\end{eqnarray*}

\subsection{Uniform electron bunch case}

For a uniform in space electron beam
\begin{eqnarray}
f_b(x,v_b) &=& \frac{1}{volume}f_v(v_b)       \nonumber\\
n_b &=& N_b/volume  \spc \spc electron \spc density     \nonumber\\
    & &       \nonumber\\
F_i &=& -4 \pi m_b n_b r_{ab}^2 c^4 \int d^3v_b \spc
     \frac{(v_a-v_b)_i}{|v_a-v_b|^3} 
  f_v(v_b) ln \lbr \frac{\beta_{ab}^2 b_{maxab}}{r_{ab}} \rbr
          \nonumber\\
\end{eqnarray}
This result for the friction force for a uniform in space electron
beam is the same as the result found using the usual theory of
electron cooling.

\subsection{Gaussian bunch case}
\begin{eqnarray}
f_b(x,v_b) &=& \frac
    {exp[-x^2/(2\sig_x^2)-y^2/(2\sig_y^2)-s^2/(2\sig_s^2)]} 
    {(2 \pi)^{3/2}\sig_x \sig_y \sig_s }
       f_{v}(v_b)       \nonumber\\
     & &       \nonumber\\
F_i &=& -4 \pi m_b N_b  r_{ab}^2 c^4 
\frac     {exp[-x^2/(2\sig_x^2)-y^2/(2\sig_y^2)-s^2/(2\sig_s^2)]} 
    {(2 \pi)^{3/2}\sig_x \sig_y \sig_s }     \nonumber\\
    & & \int d^3v_b 
     \frac{(v_a-v_b)_i}{|v_a-v_b|^3} 
  f_v(v_b) ln \lbr \frac{\beta_{ab}^2 b_{maxab}}{r_{ab}} \rbr
          \nonumber\\
\end{eqnarray}

This result can be generalized to apply to any electron bunch
ditribution that can be factored and written as
\[ f_b(x,v_b)=f_x(x)f_v(v_b)  \]
One then finds
\begin{eqnarray*}
F_i &=& -4 \pi m_b N_b r_{ab}^2 c^4 f_x(x)
    \int d^3v_b 
     \frac{(v_a-v_b)_i}{|v_a-v_b|^3} 
  f_v(v_b) ln \lbr \frac{\beta_{ab}^2 b_{maxab}}{r_{ab}} \rbr
          \nonumber\\
\end{eqnarray*}

The results for the friction force given in this paper may differ from
the usual friction force results when the electron bunch distribution
can not be factored. This may happen when the alpha funtion is not zero or 
when dispersion is present.

\section{Cooling rates for $<p_ip_j>$, due to collisions, and for
  $<x_ip_i>$}

If a horizontal dispersion is present in the cooling section , 
then the cooling rate of the emittances will also depend on the
cooling rate of $<p_x p_s>$. It will be shown that the friction force
obtained as described above  when used to track a particle sample of
the ion diStribution will give the same  cooling rate for $<p_x p_s>$ as
that found using the  IBS theory of electron cooling. Similar
statements can be made for the vertical dispersion. Thus the friction
force can be used to track a bunch of ions when dispersion is present
to find the  same emittance cooling rates as those found using the
IBS theory of electron cooling.

The friction force as defined here to give the correct cooling rates, 
due to collisions,
for $<p_i^2>, \spc i=x,y,s$ also gives the correct cooling rates for
all 6 of the moments $<p_i p_j>, \spc i,j=x,y,s$. It will also be
shown that it gives the correct  cooling rates, 
due to collisions,
for $<x_i p_i>, \spc i=x,y,s$ which is required to compute the cooling
rates of the emittances.

\section {Derivation of the friction force using intrabeam scattering results 
for electron cooling}

To derive the results for the friction force, we will first find the
cooling rates for $<p_i^2>, \spc i=x,y,s$, due to collisions, using 
the friction force. We
will then find the cooling rates for $<p_i^2>, \spc i=x,y,s$ using the
methods of IBS. Comparing these two results for the cooling rates , 
due to collisions, for 
$<p_i^2>, \spc i=x,y,s$ will give us the result for the friction force.

\subsection{Cooling rate of $<p_i^2>$ from the friction force}
Let $p_{ik}, \spc i=x,y,s$ be the components of the momentum of the
$k$th ion. Let $v_{ik}, \spc i=x,y,s$ be the components of the ion
velocity. Let $N_a$ be the number of ions in the ion bunch.
Let $F_i$ be the components of the friction force acting on the ion. If the
ions are tracked using this friction force then
\begin{eqnarray}
\frac{dp_{ik}}{dt} &=& F_i  \nonumber\\
\frac{dp_{ik}^2}{dt} &=&  2 m_a v_{ik} F_i  \nonumber\\
  & &                \nonumber\\
\frac{d<p_{ik}^2>}{dt} &=&
          \frac{1}{N_a} \sum_{k=1}^{N_a} 2 m_a  v_{ik} F_i  \nonumber\\
\frac{d<p_{ia}^2>}{dt}  &=& \int d^3x d^3v_a f_a(x,v_a) 2 m_a 
 v_{ia} F_i                                 \nonumber\\	   
\end{eqnarray}
Note that $d/dt$ here gives only the rate of change of the relevant
quantity. due to collisions between ions and electrons.

\subsection{Cooling rate of $<p_i^2>$ from the IBS theory of electron
  cooling}

Let $\delta N_a$ be the number of  ions with momentum, $p_a$ in $d^3p_a$ 
and  space coordinate $x$ in $d^3x$
which are scattered by the electrons  with momentum $p_b$ in $d^3p_b$ 
which are also in  $d^3x$, in the time interval $dt$ , into the 
solid angle $d\Omega$. In a scattering event $p_a,p_b$ change to
$p_a',p_b'$ and $q_a=p_a'-p_a$ is the change in the ion momentum.
Then $\delta N_a$ is given by, Ref.[4,5],
\begin{eqnarray}
\delta N_a &=& d\Omega \spc \sig_{ab} \spc N_a f_a(x,v_a) d^3v_a |v_a-v_b|
         \spc N_b f_b(x,v_b) d^3v_b d^3x \spc dt         \nonumber\\
\end{eqnarray}
$\sigma_{ab}$ is the scattering cross section for the scattering 
of the ions from the electrons. Using this result for $\delta N_a$ one
can find that [4,5]
\[  \]
\[  \]
\begin{eqnarray}
\delta<p_{ia}^2> &=& \int d^3v_bd^3xd^3v_a [ \spc 
               f_a(x,v_a) N_bf_b(x,v_b) |v_a-v_b|     \nonumber\\
  & &             \int d\Omega \spc \sig_{ab} \delta (p_{ia}^2) ] \spc dt
              \nonumber\\
  & &                \nonumber\\
\delta (p_{ia}^2) &=& (p_{ia}+q_{ia})^2-p_{ia}^2    \nonumber\\
                  &=& 2 p_{ia} q_{ia} + q_{ia}^2   \nonumber\\
             &=& 2 p_{ia} q_{ia} \spc dropping\;q_{ia}^2\;(see\;below)
              \nonumber\\
\int d\Omega \spc \sig_{ab} \delta (p_{ia}^2) &=& \int d\Omega \spc
              \sig_{ab} 2 p_{ia} q_{ia}              \nonumber\\ 
q_{ia} &=&  p_{ia}'-p_{ia}    \nonumber\\
\end{eqnarray}

In Eq.6, $p_{ia}$ does not depend on the scattering angles $\theta,\phi$. Let 
$d_i$ be defined as
\[d_i=\int d\Omega \spc \sig_{ab}  q_{ia} \]
$d\Omega \sig_{ab}$ is an invariant and $q_{ia}$ is a vector in 3-space
which has the same magnitude in the Rest CS and in the Center of
Mass CS (CMS). Then  $d_i$ is a vector in 3-space and can be evaluated
in the CMS.

If this integral is  evaluated in the CMS 
and the result is written in terms of tensors in 3-space then the 
result will also hold in the Rest CS.

In the CMS, we introduce a polar coordinate system $\theta,\phi$
where $\theta$ is measured relative to the direction of 
$\vec{p_a}$ 
and we assume that $\sigma_{ab}(\theta,\phi)$ is a fumction of $\theta$ only.
we can then write
\begin{eqnarray}
\vec{p_a} &=& (0,0,1)|\vec{p_a}|    \nonumber\\
\vec{p_a \; '} &=& (\sin \theta \cos \phi,\sin \theta \sin \phi,
          \cos \theta)|\vec{p_a}|    \nonumber\\
\vec{q_a} &=& (\sin \theta \cos \phi,\sin \theta \sin \phi,
          \cos \theta-1)|\vec{p_a}|    \nonumber\\
\end{eqnarray}
 In the CMS,
using Eq.7, one finds
\begin{eqnarray}
d_i &=& -2 \pi  \int d\theta sin\theta 
            (1-cos\theta) \sigma_{ab} (0,0,1) |p_a|   \nonumber\\
\end{eqnarray}   
These results for $d_i$ in the CMS can be rewritten in terms of 
tensors in 3-space. In the CMS
\[  \]
\[  v_{ia}-v_{ib}=p_{ia}/m_a-p_{ib}/m_b=p_{ia}/\mu \]
\[  p_{ia}=\mu (v_{ia}-v_{ib})  \]
\[  \]
and
\begin{eqnarray}
d_i &=& -2 \pi  \int d\theta sin\theta  
            (1-cos\theta) \sigma_{ab}   \mu (v_{ia}-v_{ib})   \nonumber\\
\end{eqnarray}   
In this form the result will also hold in the Rest CS. 

Using the above results for $\delta (p_{ia}^2)$, due to collisions, 
 and for $d_i$ and
putting them into the result for $\delta<p_{ia}^2>$ in Eq.6, one finds
\[   \]
\[  \]
\begin{eqnarray}
\delta<p_{ia}^2> &=& \int d^3xd^3v_a  f_a(x,v_a) 2 m_a v_{ia} \nonumber\\
      & & [ -2 \pi m_b \int  d^3v_b  N_b  (v_a-v_b)_i |v_a-v_b|
      f_b(x,v_b)  \nonumber\\
      & &
     (\int d\theta sin \theta (1-cos\theta) \sig_{ab}) \spc dt  ] 
              \nonumber\\
\end{eqnarray}   

\subsection{Friction force results}

Comparing the result for $\delta<p_{ia}^2>$, due to collisions, 
found here with the result
for $\delta<p_{ia}^2>$ found in section 7.1, we get the result for
the friction force
\begin{eqnarray}
F_i &=& -2 \pi m_b N_b \int d^3v_b \spc \lbr (v_a-v_b)_i |v_a-v_b| f_b(x,v_b) 
     \int d\theta sin \theta (1-cos\theta) \sig_{ab} \rbr    \nonumber\\
\end{eqnarray}   
Using for $sig_{ab}$ the results for the coulomb croos-section given in
Eq.1 one finds
\begin{eqnarray}
\sigma_{ab} &=& (\frac {r_{ab}} {\beta_{ab}^2})^2 \frac{1}{(1-cos \theta)^2} 
   \spc \spc coulomb\; cross-section\;in\; CMS            \nonumber\\ 
r_{ab} &=& \frac{Z_aZ_b e^2}{\mu c^2}     \nonumber\\
\beta_{ab} c &=& |\vec{v_a}-\vec{v_b}|   \nonumber\\
\frac{1}{\mu} &=& \frac{1}{m_a}+\frac{1}{m_b}     \nonumber\\
    & &                    \nonumber\\
F_i &=& -4 \pi m_b N_b r_{ab}^2 c^4 \int d^3v_b \spc
     \frac{(v_a-v_b)_i}{|v_a-v_b|^3} 
  f_b(x,v_b) ln \lbr \frac{\beta_{ab}^2 b_{maxab}}{r_{ab}} \rbr
          \nonumber\\
\end{eqnarray}

We can now justify dropping the $q_{ia}^2$ term in Eq.6. We will show
that $|q_a|$ is smaller than $|p_a|$ in the Rest CS by the factor
$m_b/m_a$. Thus the $q_{ia}^2$ term in Eq.6 is smaller than the
$2p_{ia}q_{ia}$  by the factor $m_b/m_a$.

$|q_a|$  has the same vaue in the CMS and in the Rest CS. In the CMS
$|q_a|$ has the magnitude of $|p_a|$ in the CMS. In Rhic, $|q_a|$ has
the magnitude of $1e-3m_b c$  while  $|p_a|$ in the Rest CS has the
magnitude of $1e-3m_a c$. Thus $|q_a|$ is smaller than $|p_a|$ in the 
Rest CS by the factor $m_b/m_a$.

\subsection{Cooling rates for $<p_ip_j>$, due to collisions, required 
when dispersion is present}

If a horizontal dispersion is present in the cooling section , 
then the cooling rate of the emittances will also depend on the
cooling rate of $<p_x p_s>$, due to collisions. It will be shown 
that the friction force
obtained as described above  when used to track a particle sample of
the ion ditribution will give the same  cooling rate, due to
collisions, for $<p_x p_s>$ as
that found using the  IBS theory of electron cooling. Similar
statements can be made for the vertical dispersion. Thus the friction
force can be used to track a bunch of ions when dispersion is present
to find the  same emittance cooling rates as those found using the
IBS theory of electron cooling.

First let us find the cooling rate of $<p_{ia}p_{ja}>$ using the
friction force. Using the same procedure as given in section 7.1 one gets
\begin{eqnarray}
\frac{dp_{ik}}{dt} &=& F_i  \nonumber\\
\frac{d(p_{ik} p_{jk})}{dt} &=&  m_a( v_{ik} F_j+v_{jk} F_i)  \nonumber\\
  & &                \nonumber\\
\frac{d<p_{ik}p_{jk}>}{dt} &=&
          \frac{1}{N_a} \sum_{k=1}^{N_a} m_a ( v_{ik} F_j+v_{jk} F_i)  \nonumber\\
\frac{d<p_{ia}p_{ja}>}{dt}  &=& \int d^3x d^3v_a f_a(x,v_a) m_a 
( v_{ia} F_j+v_{ja} F_i)                                 \nonumber\\	   
\end{eqnarray}

This result for the cooling rate of $<p_{ia}p_{ja}>$, due to collisions,
 found using our
result for the friction force will now be shown to be the same result
as that found using the IBS theory of electron cooling [5].
The cooling rate of $<p_{ia}p_{ja}>$ using the IBS theory of electron
cooling can be found using the
the same procedure as that given in section 7.2 .
\begin{eqnarray}
\delta<p_{ia}p_{ja}> &=& \int d^3v_bd^3xd^3v_a [ \spc 
               f_a(x,v_a) N_bf_b(x,v_b) |v_a-v_b|     \nonumber\\
  & &             \int d\Omega \spc \sig_{ab} \delta (p_{ia}p_{ja}) ]
              \nonumber\\
  & &                \nonumber\\
\delta (p_{ia}p_{ja}) &=& (p_{ia}+q_{ia})(p_{ja}+q_{ja})-p_{ia}p_{ja}   \nonumber\\
                  &=&  p_{ia} q_{ja}+p_{ja} q_{ia} + q_{ia}q_{ja}   \nonumber\\
             &=&  p_{ia} q_{ja}+p_{ja} q_{ia} \spc
               dropping\;q_{ia}q_{ja}
                       \nonumber\\
\int d\Omega \spc \sig_{ab} \delta (p_{ia}p_{ja}) &=& \int d\Omega \spc
              \sig_{ab} (p_{ia} q_{ja}+p_{ja} q_{ia})              \nonumber\\ 
q_a &=&  p_a'-p_a    \nonumber\\
\end{eqnarray}

Eq.14 now becomes
\begin{eqnarray}
\delta<p_{ia}p_{ja}> &=& \int d^3xd^3v_a  f_a(x,v_a) m_a      \nonumber\\
      & & ( -2 \pi m_b \int  d^3v_b   
       ( v_{ia}(v_a-v_b)_j+v_j{a}(v_a-v_b)_i) |v_a-v_b|     \nonumber\\
      & &
     N_bf_b(x,v_b ) \spc (\int d\theta sin \theta (1-cos\theta)
      \sig_{ab}) 
      \spc dt  )               \nonumber\\
\end{eqnarray}   
which, using Eq.11 for the friction force, can be written as
\begin{eqnarray}
\delta <p_{ia}p_{ja}>  &=& \int d^3x d^3v_a f_a(x,v_a) m_a 
( v_{ia} F_j+v_{ja} F_i)  \spc dt                               \nonumber\\	   
\end{eqnarray}
This is the same result as that found using the friction force, Eq.13.

\subsection{Cooling rates for $<x_ip_i>$, due to collisions.}

First let us find the cooling rate of $<x_{i}p_{ia}>$ using the
friction force. Using the same procedure as given in section 7.1 one gets
\begin{eqnarray}
\frac{dp_{ik}}{dt} &=& F_i  \nonumber\\
\frac{d(x_{ik}p_{ik})}{dt} &=&  x_{ik} F_i  \nonumber\\
  & &                \nonumber\\
\frac{d<x_{ik}p_{ik}>}{dt} &=&
          \frac{1}{N_a} \sum_{k=1}^{N_a} x_{ik} F_i  \nonumber\\
\frac{d<x_{i}p_{ia}>}{dt}  &=& \int d^3x d^3v_a f_a(x,v_a)  
 x_{i} F_i                                 \nonumber\\	   
\end{eqnarray}
Note we are finding  only the cooling rate due to collisions and
in collisions $x$ does not change.

This result for cooling rate of $<x_{i}p_{ia}>$, due to collisions,
 found using our
result for the friction force will now be shown to be the same result
as that found using the IBS theory of electron cooling [5].
The cooling rate of $<x_{i}p_{ia}>$ using the IBS theory of electron
cooling can be found using the
the same procedure as that given in section 7.2 .
\begin{eqnarray}
\delta<x_{i}p_{ia}> &=& \int d^3v_bd^3xd^3v_a [ \spc 
               f_a(x,v_a) N_bf_b(x,v_b) |v_a-v_b|     \nonumber\\
  & &             \int d\Omega \spc \sig_{ab} \delta (x_{i}p_{ia}) ]
              \nonumber\\
  & &                \nonumber\\
\delta (x_{i}p_{ia}) &=& x_{i}q_{ia}   \nonumber\\
\int d\Omega \spc \sig_{ab} \delta (p_{ia}p_{ja}) &=& \int d\Omega \spc
              \sig_{ab} x_{i}q_{ia}              \nonumber\\ 
q_{ia} &=&  p_{ia}'-p_{ia}    \nonumber\\
\end{eqnarray}
Eq.18 now becomes, using EQ.9 FOR $\int d\Omega \spc \sig_{ab}q_{ia}$
\begin{eqnarray}
\delta<x_{i}p_{ia}> &=& \int d^3xd^3v_a  f_a(x,v_a)       \nonumber\\
      & & ( -2 \pi m_b \int  d^3v_b   
       x_{i}(v_a-v_b)_i |v_a-v_b|     \nonumber\\
      & &
     N_bf_b(x,v_b ) \spc (\int d\theta sin \theta (1-cos\theta)
      \sig_{ab}) 
      \spc dt  )               \nonumber\\
\end{eqnarray}   
which, using Eq.11 for the friction force, can be written as
\begin{eqnarray}
\delta <x_{i}p_{ia}>  &=& \int d^3x d^3v_a f_a(x,v_a)  
x_{i} F_i  \spc dt                               \nonumber\\	   
\end{eqnarray}
This is the same result as that found using the friction force, Eq.17.

Thanks are due to Alexei Fedotov for his assistance in comparing the
results of the IBS treatment of electron cooling and the results found
using the friction force. 

\section*{References}

\noindent
1. A. Piwinski Proc. 9th Int. Conf. on High Energy Accelerators (1974) 405

\noindent
2. J.D. Bjorken and S.K. Mtingwa, Part. Accel.13 (1983) 115

\noindent
3. M. Martini CERN PS/84-9 (1984)

\noindent
4. G. Parzen BNL report C-A/AP/N0.150 (2004) 

     and at http://arxiv.org/ps\uu cache/physics/pdf/0405/0405019.pdf

\noindent
5. G. Parzen BNL report C-A/AP/N0.243 (2006) 

     and at http://arxiv.org/abs/physics/0609076
%

\end{document}